\documentclass{article}
\usepackage{spconf,amsmath,amssymb,graphicx}

\title{FedCPC: an Effective Federated Contrastive Learning Method for Privacy Preserving Early-Stage Alzheimer's Speech Detection}
\name{Wenqing Wei$^{1}$\thanks{*Corresponding author. Wenqing Wei and Zhengdong Yang were NICT interns during this work.}, Zhengdong Yang$^{2,4}$, Yuan Gao$^{2}$, Jiyi Li$^3$, Chenhui Chu$^2$, Shogo Okada$^{1}$, Sheng Li$^{4*}$} 

\address{
  $^1$Japan Advanced Institute of Science and Technology, Nomi, Japan\\
  $^2$Kyoto University, Kyoto, Japan\\
  $^3$University of Yamanashi, Kofu, Japan\\
  $^4$National Institute of Information and Communications Technology, Kyoto, Japan\\
}
\begin{document}
%
\maketitle
\begin{abstract}
The early-stage Alzheimer's disease (AD) detection has been considered an important field of medical studies. Like traditional machine learning methods, speech-based automatic detection also suffers from data privacy risks because the data of specific patients are exclusive to each medical institution. A common practice is to use federated learning to protect the patients' data privacy. However, its distributed learning process also causes performance reduction. To alleviate this problem while protecting user privacy, we propose a federated contrastive pre-training (FedCPC) performed before federated training for AD speech detection, which can learn a better representation from raw data and enables
different clients to share data in the pre-training and training stages. Experimental results demonstrate that the proposed methods can achieve satisfactory performance while preserving data privacy.

\end{abstract}

\noindent\textbf{Index Terms}
Alzheimer's disease speech detection, federated learning, contrastive pre-training

\section{Introduction}
\label{sec:intro}

Alzheimer's disease (AD) has been a global problem, and the number of people affected by this kind of cognitive impairment is growing. Unfortunately, this disease still cannot be perfectly cured. Therefore, early-stage AD detection methods are required. Current bio-medical diagnoses require a comprehensive examination by medical experts, which is costly and time-consuming. Compared with these biochemical methods, machine learning-based methods from easily captured spoken language signals are much more direct and efficient.
Previous works have found speech changes in fluency \cite{szatloczki2015speaking}, prosody \cite{horley2010emotional}, and rhythm \cite{lowit2006investigation,roark2011spoken} in patients with AD.
Researchers have been motivated to study AD detection using state-of-the-art technologies, such as speech recognition \cite{Tth2015AutomaticDO,Pan2020ImprovingDO,Zhou2016SpeechRI,qin2021exploiting}, speaker recognition \cite{Pappagari2020UsingSO}, natural language processing \cite{Searle2020ComparingNL,Wankerl2017AnNB, Balagopalan2020ToBO,Yuan2020DisfluenciesAF}, and multi-modeling \cite{Rohanian2020MultiModalFW}.

Although such detection would be helpful, using patient data during model training might raise concerns about privacy issues \cite{nautsch_gdpr_2019,wu_spoofing_2015,suwajanakorn2017synthesizing}. The data of specific patients are exclusive for each medical institution, which cannot support traditional machine learning methods where all the local data are uploaded to one central server for learning a global model. Under such a setting, each medical institution does not have sufficient training data for the model; in addition, different medical institutions cannot benefit from a larger training data scale, leading to significant performance degradation. 
Existing methods focus on directly anonymizing speakers' identities \cite{justin2015speaker,qian2018hidebehind, srivastava2020evaluating,fang_speaker_2019,sweeney_k-anonymity:_2002} and achieve very good results. However, these approaches are neither cheap nor time-efficient.
Federated learning \cite{mcmahan2017communication} is a machine learning technique that trains an algorithm across multiple decentralized edge devices or servers holding local data samples without the exchange of data to each other. It enables multiple clients to build a common and robust machine learning model without sharing data, thus preserving data privacy. Considering the advantages of the federated learning method, numerous works have recently used federated learning to protect user data privacy. Li et al. \cite{li2021federated} proposed ADDETECTOR, a privacy-preserving smart healthcare system, to realize low-cost AD. 
Wang et al. \cite{wang2022federated} proposed a Blockchain-based Privacy-preserving Federated Learning scheme, which can enable the verifiability of the local models while protecting data privacy.
However, federated learning is a kind of distributed learning, so the training data on each client will be smaller than in centralized learning, which leads to performance degradation, especially for small dataset tasks. 

Several substantial works have recently demonstrated that self-supervised representations are highly successful in downstream speech and language processing tasks through feature-based speech representation extraction or fine-tuning as part of the downstream model \cite{jaiswal2020survey}. Oord et al. \cite{oord2018representation} proposed contrastive predictive coding (CPC) that seeks to group samples that are alike while keeping samples that are different from one another apart from representation learning, which can accurately represent the data. Baevski et al. \cite{baevski2021unsupervised} used adversarial training to train an unsupervised speech recognition model using the representations of the unlabeled speech audio data and the unlabeled phonemicized text data.

Inspired by these works, this paper proposes using federated contrastive learning to protect patient data privacy and enhance the model's performance for each client by enabling data sharing while preserving privacy during the pre-training and training stages. 
Specifically, each medical institution could be regarded as a client. The clients locally train an independent contrastive predictive coding pre-training model and then upload the model to a central server. 
On the server side, the multiple models are then stacked up as a global model with federated averaging and hierarchical optimization, which cannot backtrack the parameters of individual models. After completing the Federated contrastive (FedCPC) pre-training process, we apply the FedCPC pre-training model as a feature extractor in each client to detect the AD speech with federated learning. Finally, this global model is sent to each client to benefit from more extensive data and guarantee strong anonymity and privacy.

\section{Methods}
\label{sec:propose}
\begin{figure*}[!t]
  \setlength{\abovecaptionskip}{10pt}
  \centering
  \includegraphics[width=\textwidth]{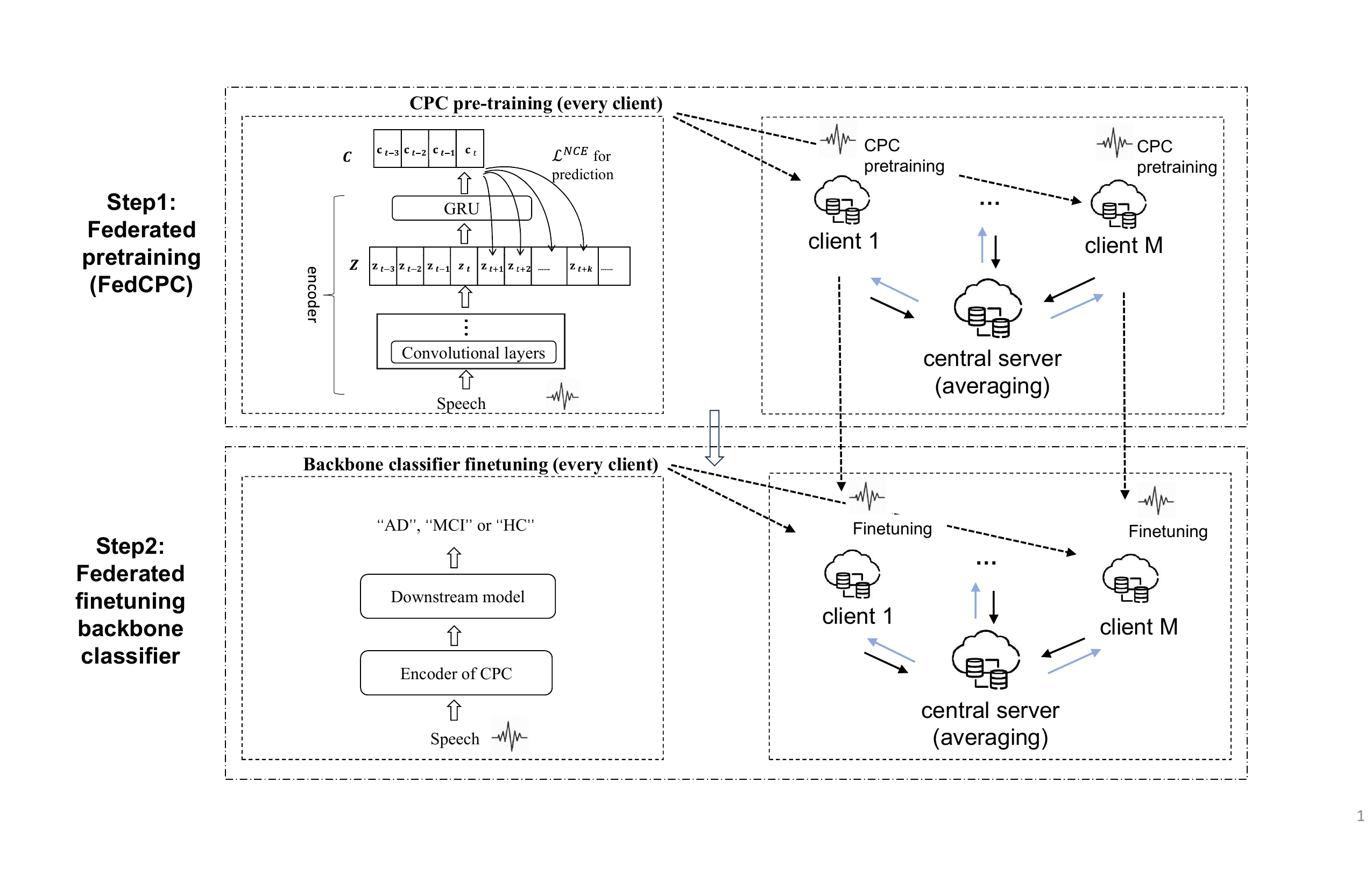}
  \caption{The flowchart of the proposed method. The algorithm on the left side of the diagram will be applied to each client on the right side. }
  \label{fig:proposed}
\end{figure*}
Figure \ref{fig:proposed} demonstrates the flowchart of the proposed method. Our method is divided into two steps. Step 1 is the proposed Federated CPC pre-training (FedCPC). The pre-training model for learning the representations of the speech data is CPC, and Federated learning is utilized for training the CPC model among the central server and the clients. After Federated CPC pre-training, Step 2 loads the pre-trained model and fine-tunes the classifier for the downstream Alzheimer's disease detection task. The details of the CPC Model and the Federated Learning module are described as follows:

\subsection{Contrastive Predictive Coding Pre-training}

Self-supervised learning obtains supervisory signals from the data, which provides supervisory signals beneficial for downstream tasks. CPC \cite{oord2018representation} is a self-supervised learning approach, which involves predicting the future time step of data according to the context vector derived from past data. Its goal is to learn representations that allow long-term prediction of future time steps by maximizing mutual information between representations and predictions. Moreover, CPC captures high-level information from a signal (for example, global structures such as phonemes in speech  \cite{oord2018representation}). This work adopts the CPC method for unsupervised representation pre-training of Alzheimer's speech.

 A convolutional encoder network produces a sequence of representation $Z$ of a raw audio waveform. Subsequently, a recurrent context network summarizes the past information in the vector embedding sequence and produces corresponding contextual $\mathbf{C} = \{\mathbf{c}_1, \mathbf{c}_2,...,\mathbf{c}_t\}$ representations for audio.
The CPC pre-training model is trained to predict the future latent representation $\mathbf{z}_{t+k}$ ($k$ is the predicted step) using the context-aware representation $\mathbf{c}_t$ at the $t$-th timestep of speech. At each step $t$, we adopt a contrastive estimation-based InfoNCE \cite{oord2018representation} to maximize the mutual information lower bound between contextual representations $\mathbf{c}_t$ and future latent representations $\mathbf{z}_{t+k}$. 
Here, given a set $Z = \{\mathbf{z}_1,\mathbf{z}_2,...,\mathbf{z}_N\}$ N random samples which contains one positive sample from $p(\mathbf{z}_{t+k} | \mathbf{c}_t)$ and $N-1$ negative samples from ``noise" distribution $p(\mathbf{z}_{t+k})$ are drawn for optimizing the loss.
And The formula is as follows:
\begin{align}
    \mathcal{L}^{NCE}_{tk} = -\mathbb{E} \left[ \log \frac{\exp(\mathbf{c}^{T}_{t} \mathbf{W}_{k} \mathbf{z}_{t+k}))}{\frac{1}{N} \sum_{\Tilde{\mathbf{z}} \in {Z}} \exp(\mathbf{c}^{T}_{t}\mathbf{W}_{k} \mathbf{\Tilde{\mathbf{z}}})} \right]
\end{align}

\subsection{Federated Learning}
With the advent of the General Data Protection Regulation and increasing privacy concerns, sharing real-world AD speech data is facing significant challenges \cite{nautsch_gdpr_2019}. Speech data contain the speaker's identifiable information represented as voiceprint used in many authentication systems \cite{boles_voice_2017, wechat2015}. Exposing an individual's voiceprint may cause security risks \cite{wu_spoofing_2015, suwajanakorn2017synthesizing} to voice authentication systems. Several methods have been proposed to anonymize speakers' identities using state-of-the-art technologies, such as
speech synthesis \cite{justin2015speaker}, voice conversion \cite{qian2018hidebehind, srivastava2020evaluating}, speaker embeddings \cite{fang_speaker_2019}, k-anonymity model \cite{sweeney_k-anonymity:_2002}. However, these approaches are neither cost nor time-efficient and, therefore, do not meet the demands for big data on a global scale.

Federated learning proposes a solution to privacy-preserving large-scale machine learning by first training models locally on an individual client device and then aggregating the updates of the local models on the central server. There are several proposed algorithms for solving the Federated Optimization problem. One of the most promising algorithms is federated averaging (FedAvg) \cite{mcmahan2017communication}. The algorithm provides privacy by design and personalizes models to individual users. It also claims to be more resource-efficient in terms of communication rounds.

This algorithm, firstly proposed in \cite{mcmahan2017communication}, relies on the stochastic gradient descent (SGD) optimization method since most of the most successful deep learning works were based on this. The available clients locally compute their average gradient on their local data at the parameters of the current model $w_s$, where $s$ identifies the federated round. The central server aggregates these gradients and applies the update $w_{s+1}\leftarrow w_s - \eta\sum_{m=1}^M\frac{n_m}{n}g^{m}_s$, where $g^{m}_s=\nabla F_m(w_s)$ is the gradient of the client $m$ in $s_{th}$ federated round, $\eta$ is the learning rate, $n_m$ is the number of samples at the client $m$, $n$ is the total number of samples (sum over all the available clients), $M$ is the number of the clients. Equivalently, the update can be given by $w_{s+1}\leftarrow \sum_{m=1}^M\frac{n_m}{n}w_{s+1}^m$, where $w_{s+1}^m\leftarrow w_s - \eta g^{m}_s$, $\forall m$. Finally, every client takes a complete gradient descent step, while the server only takes the weighted average of the resulting models. 

\subsection{Overview of proposal model}
As described above, federated learning is used in both steps of our proposed FedCPC-based model. For the FedPCPC pre-training model proposed in step 1, $w_s$ contains all the parameters of the FedPCPC pre-training model in the $s_{th}$ federated round. These parameters are shared through federated learning, which enables different clients to learn speech representation. For the downstream part of the FedCPC-based model of step 2, $w_s$ contains the CPC pre-training model's encoder part and the downstream model's parameters in $s_{th}$ federated round.

\section{Experiments}
\label{sec:exp}

\subsection{Data Description}
\label{ssec:data}
The 2021 NCMMSC AD Recognition Challenge provides the dataset. Since we joined the short speech track, we segmented all the long sentences from the data provided by the organizers into short speech clips of 6 seconds. As shown in Table~\ref{tab:data}, we selected a 7.16-hour speech set from 39 male speakers and 54 female speakers as the training set (Training) and a 0.67-hour speech set from 15 male and 15 female speakers as the development set (Development). The test set is the official short speech track test set with a 1.92-hour speech set (Testing). All the sentences have labels in three kinds: Alzheimer's disease (AD), mild cognitive impairment (MCI), and health common (HC).

\subsection{System Implementation}

We construct the following models both with conventional centralized training and federated learning, and their implementations are described as follows:

\subsubsection{Baseline}

\begin{enumerate}
	\item A 20-dimensional Mel frequency cepstral coefficients (MFCC) vector extracted with a 25ms window and 10ms frameshift is employed as the input feature of each frame. The CNN (convolutional neural network)-based system\footnote{Available at https://github.com/THUsatlab/AD2021} has 259 $\times$ 20 nodes as the input layer, three nodes as output, and five convolutional layers with one max-pooling layer following every layer. The convolutional filters of each layer are sequentially arranged as 32, 32, 32, 64, and 128. Finally, the last two layers are fully connected (FC) with 256 nodes before softmax output. 
	
	\item CNN is mainly for learning local features. To consider contextual dependencies from local features, we construct a CNN-LSTM-based model. Two LSTM layers are put on top of the CNN-based model, and this model also has two FC layers before softmax output. The network configuration of the CNN and FC layers is the same as the CNN-based system. 
	
	\item Moreover, to compare with other pre-trained models, we adopt AST\footnote{Available athttps://github.com/YuanGongND/ast} model \cite{gong21b_interspeech}. It is transformer-based
model for audio classification with a larger database with a simple architecture and superior performance. For this reason, we use the AST model to compare with our CPC-base model.

\end{enumerate}

\begin{table}[!t]
  \centering
  \caption{Data Descriptions}
 \scalebox{1.0}{ 
  \begin{tabular}{l|c|c|c}
    \hline
    \multicolumn{1}{c|}{\textbf{}} & \multicolumn{1}{c|}{\textbf{\#speakers}} & \multicolumn{1}{c|}{\textbf{\#utterance}} & \multicolumn{1}{c}{\textbf{\#hour}} \\
    \multicolumn{1}{c|}{\textbf{}} & \multicolumn{1}{c|}{\textbf{(Male/Female)}} & \multicolumn{1}{c|}{\textbf{(HC/MCI/AD)}} & \multicolumn{1}{c}{\textbf{}} \\
\hline
Training & $39$ / $54$  & $1712$ / $1380$ / $1208$  &  $7.16$ \\
Develop. & $15$ / $15$ & $114$ /  $145$ /   $138$ & $0.67$ \\
Testing & / & $432$ /  $378$ /   $343$  & $1.92$ \\
\hline
  \end{tabular}}
  \label{tab:data}
\end{table}

\begin{table}[!t]
  \centering
  \caption{Major Experimental Settings}
 
\begin{tabular}{l|l|l}
\hline
\textbf{Training settings} 
& GPUs (3090) & 1 \\
& Batch-size & 128 \\
& Epochs & 100 \\
& Steps &  34 \\
& Optimizer & Adam \\
& Valid & Dev. Set \\
\hline
\textbf{Federated settings} &  Federated round & 50 \\
& Local epochs & 4 \\
& Weighting Strategies & FedAvg \\
& \#Clients & 3 \\
\hline
\end{tabular}
\label{tab:expsetting}
\end{table}

\subsubsection{Self-Supervised Model}
\textbf{Pre-training model setup:} The implementation of CPC pre-training model  is similar to \cite{oord2018representation}. Each training iteration randomly extracts a segment of about 20,480 frames from every utterance as the encoder's input. The encoder comprises five 1-dimensional CNN layers and a single-layer gated recurrent unit (GRU). In detail, each of the five layers has the same down-sampling rate of 1/160 to get the same frame rate and the same settings of the filter size, strides, and paddings ([10, 8, 4, 4, 4],  [5, 4, 2, 2, 2] and [3, 2, 1, 1, 1]). All five layers have 512 hidden units. Moreover, the GRU layer is employed as the sequence model with  256 hidden units. Every frame of GRU output is used to predict the context $\mathbf{c}$ (12 future frames). Adam optimizer trains the model with a learning rate of 2e-4 and a minibatch size 8. 

\noindent\textbf{Backbone classifier network:} As AST is a pre-trained model used to compare our proposed method, we only constructed CNN and CNN-LSTM models as downstream models in this work. The training parameters are the same as the baseline and the scheme above.

We used the open-sourced federated learning framework Flower \cite{beutel2022flower} and PyTorch version-1.9.1\footnote{https://pytorch.org/} for all our experiments. Table \ref{tab:expsetting} lists the training and test settings. Moreover, speakers of each client's data are unique to ensure the independence of speakers from different clients in federated learning.

\section{Reuslts and Discussions }
\label{ssec:results}
\subsection{Main results }

\begin{figure}[!t]
  \setlength{\abovecaptionskip}{10pt}
  \centering  \includegraphics[width=0.46\textwidth]{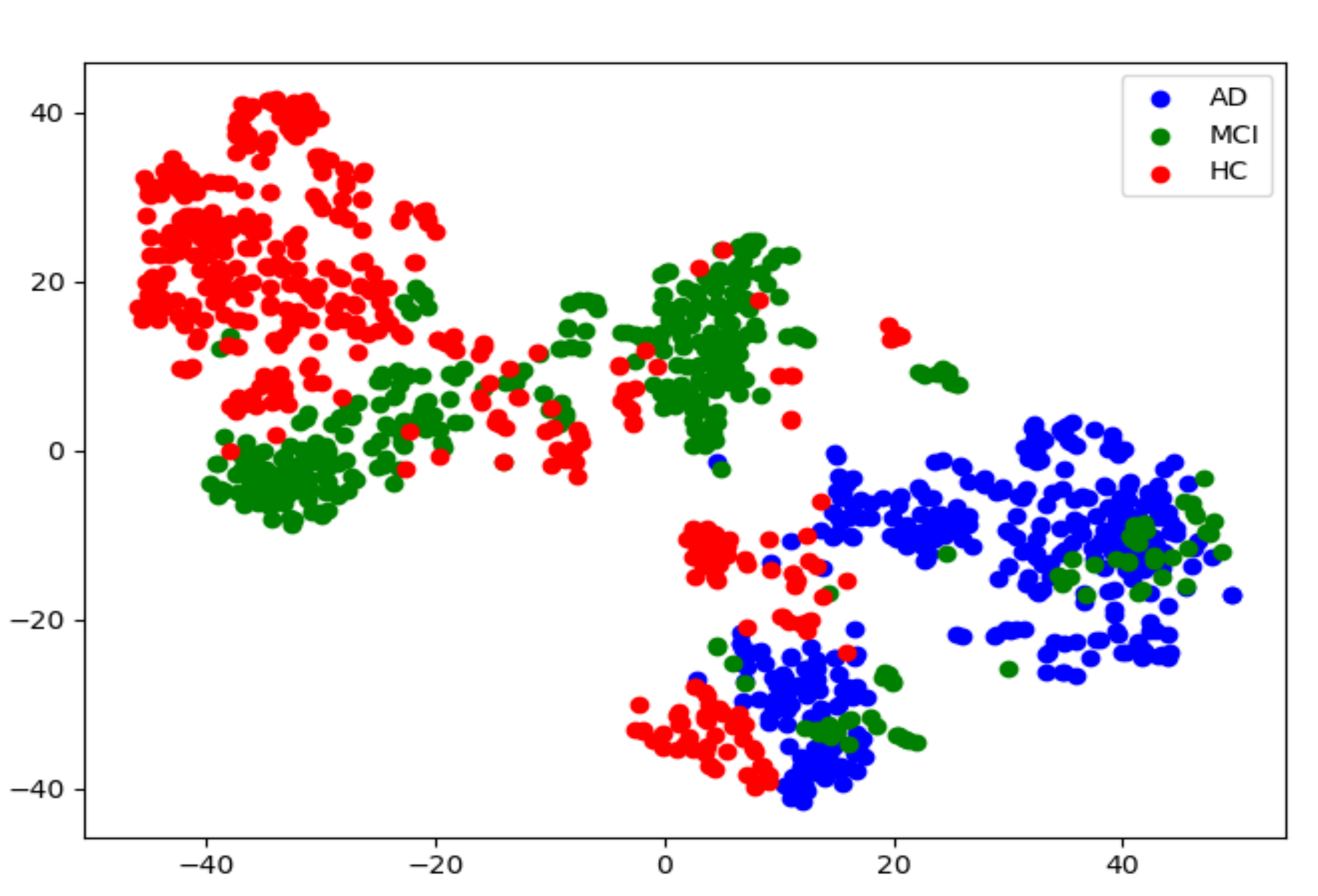}
  \caption{t-SNE visualization of audio with CPC pre-training.}
  \label{fig:AD_tsne_ssl}
\end{figure}

Using the centralized learning scenario, we first investigate whether the CPC pre-training model can improve performance. Table \ref{tab:result-centralized-learning} shows the experimental results of the test set of centralized learning. Compared with the CNN model, the CNN-LSTM model shows better performances in both machine learning paradigms, especially outperforming the CNN system on precision, recall, and F1-score by 3.8$\%$, 4.2$\%$, and 5.0$\%$ in centralized learning. Meanwhile, using CPC pre-training models leads to better performance than MFCC features. As shown in Table \ref{tab:result-centralized-learning}, the macro F1-score achieved 8.5$\%$, 3.6$\%$ improvements in centralized learning by comparing CNN and CNN-LSTM models, and CPC-CNN almost got the same F1-score with CPC-LSTM-CNN.
Moreover, the paper \cite{qin2021exploiting} utilizes the Wav2vec speech recognition pre-training model to predict Alzheimer's Disease on the same databases. In contrast, our proposed method leverages the CPC pre-training model without federated learning, specifically the CPC-CNN-LSTM model, and achieves a superior result (F1: 78.8\%) compared to the best result obtained by Wav2vec2.0\_3-2 (F1: 77.2$\%$) in \cite{qin2021exploiting} on short audio tracks. These findings provide evidence that the CPC pre-training model captures structural information embedded in raw audio signals, enhancing AD detection performance. Figure \ref{fig:AD_tsne_ssl} illustrates the visualization of representations learned by the pre-training model, indicating specific properties that can be utilized for clustering in this task.

\begin{table}[!t]
\small
  \centering
  \caption{Evaluation with centralized learning.}
  \scalebox{1.0}{
  \begin{tabular}{l|c|c|c}
    \hline
    \multicolumn{1}{c|}{\textbf{}} & \multicolumn{1}{c|}{\textbf{Precision(\%)}} & \multicolumn{1}{c|}{\textbf{Recall(\%)}} & \multicolumn{1}{c}{\textbf{F1(\%)}} \\
    \hline
CNN & 	$71.3$ &	$ 71.1 $  &	$70.2$  \\
\hline
CNN-LSTM  & $75.1$ &	$75.3$ &	$75.2$ \\
\hline
CPC-CNN  & 	$80.3$&	$78.4$ &  	$78.7$ \\
\hline
CPC-CNN-LSTM & $\bf{84.7}$ & $\bf{78.4}$ &	$\bf{78.8}$\\
\hline
AST & $75.2$ &	$74.3$  &	$75.5$\\
\hline
Wav2vec2.0\_3-2 \cite{qin2021exploiting} & $77.9$ &	$77.2$  &	$77.2$\\

\hline
  \end{tabular}}
  \label{tab:result-centralized-learning}
\end{table}

 \begin{table}[!th]
 \small
   \centering
   \caption{Evaluation on the testing set with federated learning.}
    \resizebox{\linewidth}{16mm}{
   \begin{tabular}{@{}l|c|c|c|c@{}}
     \hline
     \multicolumn{1}{c|}{\textbf{Method}} & \textbf{Downstream Model} &
     \multicolumn{1}{c|}{\textbf{Precision(\%)}} & \multicolumn{1}{c|}{\textbf{Recall(\%)}} & \multicolumn{1}{c}{\textbf{F1(\%)}} \\
     \hline
 Fed-AST & - & $73.7$ & $72.5$  & $73.5$\\
 \hline
 Fed & CNN & $72.8$  & $70.2$ &  $68.2$  \\
 CPC-Client & CNN & $77.0$ &	$71.8$  &	$72.3$\\
 FedCPC (Our) & CNN & ${\bf79.3}$ &	${\bf74.3}$  &	${\bf74.8}$\\
 \hline
 Fed & CNN-LSTM & $72.0$ & $72.1$ & $71.4$ \\
 CPC-Client & CNN-LSTM & $76.4$ &	$72.9$  &	$73.5$\\
 FedCPC (Our) & CNN-LSTM & ${\bf79.5}$ &	${\bf74.7}$  &	${\bf75.3}$\\
 \hline
       \end{tabular}
       }
   \label{tab:result-Federated-learning}
 \end{table}

\begin{figure}[!t]
  \setlength{\abovecaptionskip}{10pt}
  \centering 
  \includegraphics[width=0.47\textwidth]{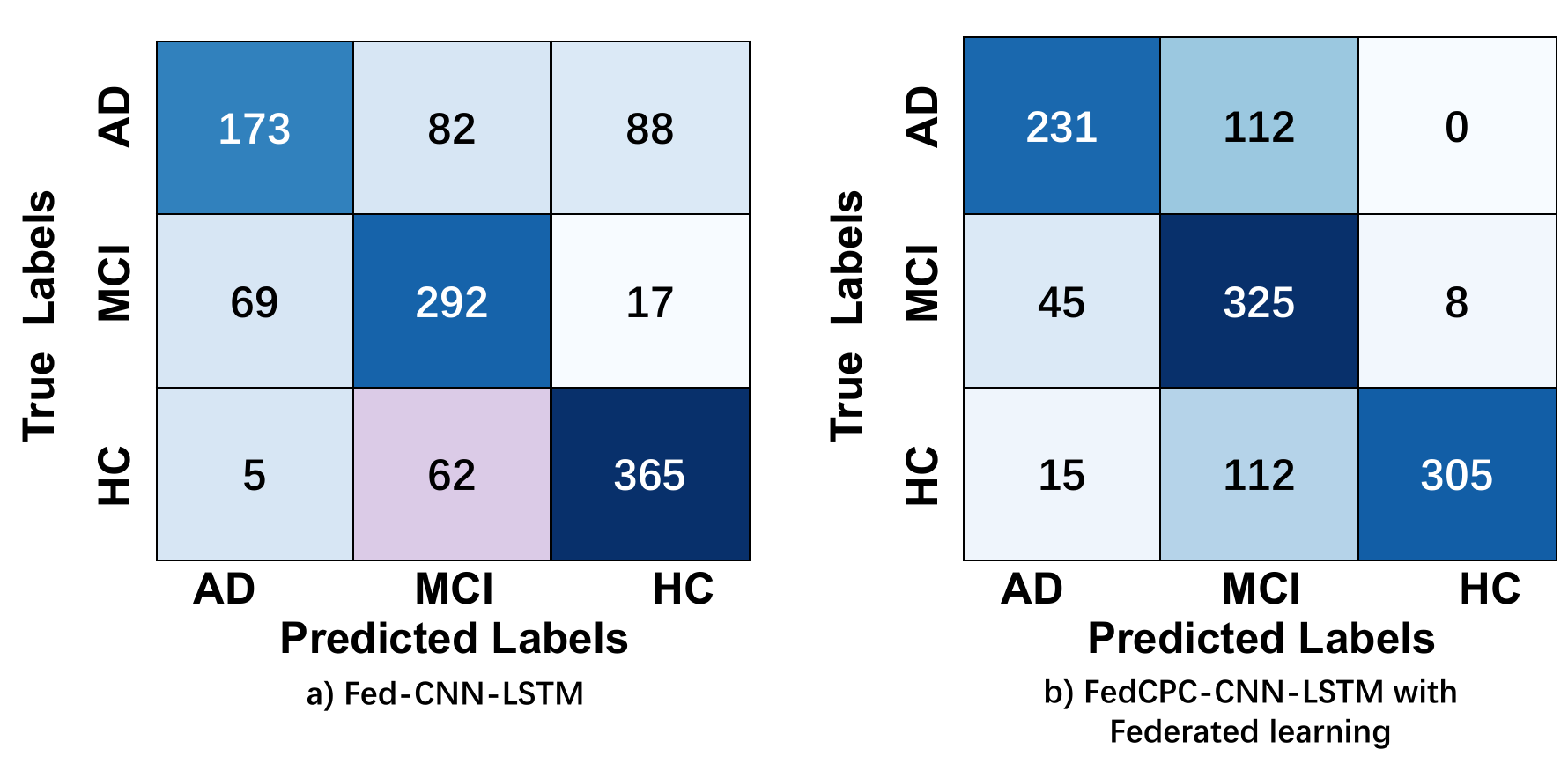}
  \caption{Confusion matrix of federated learning (a) Fed-CNN-LSTM, (b) FedCPC-CNN-LSTM.}
  \label{fig:confusion}
\end{figure}

We then verify the performance of our proposed FedCPC-based models. Table \ref{tab:result-Federated-learning} demonstrates the results of federated learning. The CPC-Client-based model used the pre-training models trained by different clients using data unique to each client. The FedCPC-based model used the pre-training models trained with federated learning. As shown in Table \ref{tab:result-Federated-learning}, the FedCPC-based model outperforms the CPC-Client model. Our proposed FedCPC-CNN-LSTM model gets the best result in federated learning. Compared with the CNN-LSTM model, the FedCPC-CNN-LSTM model significantly improves from 71.4$\%$ to 75.3 $\%$ in the F1-score. For comparison, we also exploit a pre-training AST model on this task and notice that FedCPC-CNN-LSTM outperforms it.
Meanwhile, we check the results for different categories by using confusion matrices to explore the reasons for the improvement. The confusion matrices for Fed-CNN-LSTM and FedCPC-CNN-LSTM with features computed on the test set, as depicted in Figure \ref{fig:confusion}. We found that Fed-CNN-LSTM tends to classify AD into the MCI and HC category incorrectly, while FedCPC-CNN-LSTM tends to classify AD into the MCI. It is consistent with the distribution of audio in Figure \ref{fig:AD_tsne_ssl}. Moreover, FedCPC-CNN-LSTM performs better on AD and MCI.

\subsection{Further Discussions}
With the results in Tables \ref{tab:result-centralized-learning} and \ref{tab:result-Federated-learning}, we found that the results of the federated-learning-based models have decreased compared with the traditional centralized learning results. The recall and F1-score of the CNN-LSTM system decreased by 3.1$\%$, 3.2$\%$, and 3.8$\%$ after adopting federated learning. Moreover, the FedCPC-CNN-LSTM dropped by 1.8$\%$ and 3.4$\%$ in the recall and F1-score after adopting federated learning, respectively. This result is expected regarding each client side's limited data and partial observations. This shows the inherent limitations of federated learning, which are worth investigating. 
The AST cannot outperform the pre-trained small models in the experiments above. Considering its high cost, it is not applicable for memory-restricted clients.

\section{Conclusion}
\label{sec:conclude}

This paper uses federated learning to train models to detect AD in speech and protect privacy in the user's voice. Compared with centralized learning, federated learning is a distributed training mode, leading to performance degradation, especially in small databases.
To address this issue, we proposed the FedCPC-based pre-training method, which enables data sharing while preserving privacy during the pre-training and training stage. The experimental evaluation revealed that our proposed approach effectively preserves privacy while maintaining competitive performance compared to non-privacy-preserving methods.

\section{Acknowledgement}
\label{sec:ack}
This work was supported by JSPS KAKENHI Grant Numbers JP23K11227, JP23H03454, and NICT tenure-track funding. 

\bibliographystyle{IEEEbib}\footnotesize
\bibliography{voice-privacy}

\end{document}